\documentclass[letterpaper]{article}

\usepackage[accepted]{icml2026}

\usepackage{amsmath}
\usepackage{amssymb}
\usepackage{mathtools}
\usepackage{booktabs}
\usepackage{graphicx}
\usepackage{multirow}
\usepackage{enumitem}

\usepackage[utf8]{inputenc}
\usepackage[T1]{fontenc}
\PassOptionsToPackage{hyphens}{url}
\usepackage{hyperref}
\usepackage[capitalize,noabbrev]{cleveref}
\usepackage{microtype}
\usepackage{eurosym}
\usepackage{listings}
\crefname{lstlisting}{Listing}{Listings}
\Crefname{lstlisting}{Listing}{Listings}

\definecolor{regokw}{RGB}{0,90,160}
\definecolor{regocm}{RGB}{0,128,80}
\definecolor{regostr}{RGB}{160,60,20}
\lstdefinelanguage{rego}{
  keywords={package, import, default, allow, deny, violation, some, not, with, as, in, if, else, contains},
  morekeywords=[2]{input, data, count, sprintf, object, sum},
  sensitive=true,
  morecomment=[l]{\#},
  morestring=[b]",
}
\icmltitlerunning{Governance-as-Code: EU AI Act Compliance Pipelines for Generative AI}

\begin{document}

\twocolumn[
\icmltitle{Governance-as-Code: Translating EU AI Act Technical Requirements\\into Executable Compliance Pipelines for Generative AI Systems}

\icmlsetsymbol{equal}{*}

\begin{icmlauthorlist}
\icmlauthor{Rudrendu Kumar Paul}{bu}
\icmlauthor{Sourav Nandy}{ut}
\end{icmlauthorlist}

\icmlaffiliation{bu}{Boston University, Boston, MA, USA}
\icmlaffiliation{ut}{University of Texas at Austin, Austin, TX, USA}

\icmlcorrespondingauthor{Rudrendu Kumar Paul}{rudrendupaul2022@gmail.com}
\icmlcorrespondingauthor{Sourav Nandy}{sourav.nandy@gmail.com}

\vskip 0.3in
]

\printAffiliationsAndNotice{Accepted at the AI4Law Workshop, ICML 2026.}

\begin{abstract}
The EU AI Act (Regulation 2024/1689) imposes technical obligations on high-risk AI providers, yet Articles~8--15 were drafted for predictive AI and leave seven technical gaps when applied to generative systems, spanning non-deterministic data governance, training-data provenance, continuous conformity, human oversight, open-ended robustness, emergent risk, and generative fairness. We deliver \textsc{Governance-as-Code} (GaC), a framework of 43 machine-checkable acceptance criteria across six compliance modules that run in a CI/CD pipeline and emit Article-indexed audit evidence, and we show the actual Rego policy code rather than merely describing it. Our central commitment is that the Act's open-textured standards (``appropriate levels,'' ``possible biases'') become declared, auditable numbers: robustness thresholds are derived from the provider's documented baseline and a state-of-the-art floor, and framing bias is collapsed into eight measurable proxies tested by counterfactual demographic probing. We also correct who owes what, since under Article~25 and Chapter~V a downstream deployer relies on the upstream provider's Article~53 training-data summary and documents only the layers it controls, so GaC verifies that summary rather than demanding per-sample documentation the deployer never had. We validate on two enterprise deployments, a high-risk advisory chatbot and a limited-risk content generator, benchmarking against a manual expert audit rather than documentation artifacts that were never designed to enforce compliance. GaC reproduces all of the manual audit's findings, including three penalty-triggering violations, while cutting audit labor by roughly 75\%.
\end{abstract}

\section{Introduction}

On August~2, 2026, the high-risk provisions of the EU AI Act take full effect. Organizations deploying AI in hiring, credit scoring, law enforcement, and critical infrastructure face penalties of up to \euro{}15 million or 3\% of global annual turnover for non-compliance with high-risk obligations \citep{EU_AI_Act_2024}. With the deadline close, many engineering teams still cannot answer a basic question: does our generative AI system actually satisfy Articles~8 through~15?

The problem is not a lack of awareness. The Act's risk-based architecture has been analyzed extensively \citep{Veale_2021, Novelli_2024, Smuha_2021}. Documentation frameworks exist \citep{Mitchell_2019, Gebru_2021, Arnold_2019}, governance management standards exist \citep{NIST_AI_RMF_2023, ISO_42001_2023}, and recent work has begun mapping legal requirements to verification activities \citep{Buscemi_2025}. What is missing is the last mile: turning regulatory text into machine-checkable acceptance criteria that run automatically in a deployment pipeline, with the operationalizing choices made explicit rather than hidden inside an auditor's head.

Consider a concrete failure. An organization deploys a retrieval-augmented advisory chatbot classified as high-risk under Annex~III, Area~5(b) (creditworthiness assessment). Article~9 requires a risk management system that operates ``throughout the entire lifecycle.'' The team runs a conformity assessment at deployment. Three weeks later the retrieval index is rebuilt with updated domain documents, shifting the embedding distribution; the system's hallucination rate on domain questions climbs from 4\% to 9\%. No re-assessment fires, because the change does not meet Article~43's ``substantial modification'' threshold: the model weights are unchanged, only the retrieval corpus shifted. The risk management system, built around a point-in-time audit, is blind precisely where generative systems drift.

This gap between regulation and implementation recurs across Articles~8--15. Each instance traces to an assumption in the legal text that holds for predictive ML but breaks for generative systems. The Act's requirements presume bounded output spaces, deterministic inference, tractable data provenance, and human-interpretable decision boundaries. Generative systems violate all four at once.

A second, subtler problem sits underneath the first. Even where a requirement \emph{can} be operationalized, doing so forces a choice that the legal text leaves open. Article~15 asks for ``appropriate levels'' of robustness; turning that into a pass/fail gate means picking a number. Article~10 asks teams to examine data for ``possible biases''; checking that for an open-ended generator means choosing a measurable proxy for ``bias.'' If those choices are buried as constants in a script, the compliance claim is no more auditable than a PDF. Our central design commitment is that every such choice is \emph{declared}, justified against the Act's language, and recorded as evidence.

\textbf{Contributions.}
\begin{enumerate}[leftmargin=*,itemsep=0pt,parsep=0pt,topsep=2pt]
\item A \textbf{systematic analysis of seven technical gaps} in applying Articles~8--15 to generative AI, grounded in the regulation's text, its allocation of duties across the value chain (Articles~25 and~53), and its implementation timeline (\cref{sec:gaps}).
\item \textbf{\textsc{Governance-as-Code}}, a framework of 43 machine-checkable acceptance criteria in six modules. We show the \emph{actual} policy code (\cref{sec:framework}) and make explicit how each vague standard becomes a declared, auditable threshold (\cref{sec:operationalize}).
\item \textbf{Empirical validation} on two enterprise case studies, benchmarked against a manual expert audit rather than documentation artifacts, with full reporting of what each violation means and how it was measured (\cref{sec:evaluation}).
\end{enumerate}

\section{Background and Related Work}
\label{sec:background}

\subsection{The Act's technical requirements and who owes them}

Regulation~2024/1689 establishes a risk-based classification in which high-risk systems (Annex~III) must satisfy Articles~8--15 \citep{EU_AI_Act_2024}. Article~8 mandates compliance ``taking into account the generally acknowledged state of the art.'' Articles~9--15 then specify risk management (Art.~9), data governance (Art.~10), technical documentation (Art.~11), record-keeping (Art.~12), transparency (Art.~13), human oversight (Art.~14), and accuracy, robustness, and cybersecurity (Art.~15).

A point that prior technical work often glosses, and that materially shapes any compliance pipeline, is that these duties are split across the value chain. General-purpose AI (GPAI) models carry their own obligations under Chapter~V, Articles~51--55. In particular, Article~53(1)(d) requires the upstream model provider to publish a ``sufficiently detailed summary'' of training content, using the Commission's mandated template \citep{EU_AI_Act_Art53, GPAI_CoP_2025}. Article~25 then governs how responsibility flows downstream: a deployer that fine-tunes or substantially modifies a third-party model can be reclassified as a provider, but for the upstream model itself the deployer's duty is largely to \emph{rely on and pass through} the documentation the provider supplies under Article~53, secured by the written agreement Article~25(4) anticipates \citep{EU_AI_Act_Art25}. The deploying organization does not, in the common case, control or owe documentation for the foundation model's training corpus. It owes documentation for the layers it does control: fine-tuning data, alignment data, and the retrieval corpus. We take this division seriously throughout, because a framework that demands per-sample documentation of a web-scale corpus from a downstream deployer is asking for something the Act itself does not require.

\subsection{Existing governance approaches}

\textbf{Documentation artifacts.} Model Cards \citep{Mitchell_2019}, Datasheets \citep{Gebru_2021}, and IBM AI FactSheets \citep{Arnold_2019} provide structured transparency. \citet{Golpayegani_2024} extended this with AI Cards, a machine-readable documentation framework using Semantic Web technologies aligned with the Act. These artifacts capture system properties at a point in time. They do not execute checks, enforce constraints, or detect when documented properties diverge from runtime behavior. A Model Card reporting a 3\% hallucination rate stays unchanged when production climbs to 9\%. This is a difference in \emph{purpose}, not a deficiency: documentation artifacts were designed to communicate, not to gate a deployment. That distinction matters for how we benchmark (\cref{sec:evaluation}).

\textbf{Management-system standards.} The NIST AI Risk Management Framework \citep{NIST_AI_RMF_2023} and ISO/IEC~42001 \citep{ISO_42001_2023} specify organizational processes for governing AI risk (Govern, Map, Measure, Manage, in NIST's case). They define \emph{what} an organization should do but, by design, leave the technical realization open. GaC is complementary: it is one concrete, code-level realization of the ``Measure'' and ``Manage'' functions for the specific obligations in Articles~8--15.

\textbf{Compliance mapping frameworks.} \citet{Buscemi_2025} decomposed high-risk requirements into verification activities mapped to lifecycle stages. \citet{BuenoMomcilovic_2024} combined ontologies, assurance cases, and factsheets to argue robustness compliance for LLMs. Both clarify \emph{what} must be verified but stop short of \emph{how to automate} that verification in a pipeline.

\textbf{Runtime enforcement.} \citet{Gaurav_2025} proposed Governance-as-a-Service, a multi-agent layer that intercepts agent outputs at runtime using trust scoring and graduated interventions. \citet{Kaptein_2026} formalized runtime governance as deterministic functions over execution paths. Both address runtime behavior but are not structured around the Act's articles and do not produce the article-indexed audit evidence a conformity assessment needs.

\textbf{Generative bias and robustness measurement.} Because two of our gaps (G5, G7) turn on measurement, we draw on the open-ended generation literature. BOLD \citep{Dhamala_2021} and HolisticBias \citep{Smith_2022} show that bias in free-form text can be measured through counterfactual prompt sets and named output metrics rather than through a fixed classification target; we adapt that methodology into a checkable proxy (\cref{sec:operationalize}). For robustness, HELM \citep{Liang_2023} demonstrates large variation across task types, and the indirect prompt-injection results of \citet{Greshake_2023} supply the threat model our robustness module tests against.

\subsection{The missing piece}

\citet{Marino_2026} proposed computational compliance as a research program for integrating algorithmic approaches across the AI lifecycle. We share the diagnosis but note the work remains a blueprint: it defines objectives without delivering article-specific acceptance criteria. \citet{Nolte_2025} identified legal gaps in the Act's robustness and cybersecurity provisions, and \citet{Deckenbrunnen_2026} proposed technical sandboxes as micro-foundations for regulatory learning. Neither translates these insights into executable checks. Where prior work maps, documents, or enforces, we \emph{compile}: we translate requirements into acceptance tests a pipeline runs before every deployment, producing machine-readable, article-indexed evidence.

\section{Seven Technical Gaps}
\label{sec:gaps}

\textbf{How we derived the gaps.} We did not aim to enumerate every conceivable mismatch; we aimed for the gaps that are (a) grounded in a specific obligation in Articles~8--15, (b) triggered by a concrete technical property of generative systems, and (c) addressable by a pipeline check. We worked article by article, asking for each: can an engineering team implement this requirement \emph{as written} for an LLM-based system? Where the answer was no, we recorded the breaking assumption. The seven gaps are therefore a deliberately operational set, not a claim of exhaustiveness. They are not fully orthogonal, and we do not claim they are. Article~9 appears in two gaps (G3 continuous conformity, G6 emergent risk) because it carries two distinct assumptions; Article~10 appears in G1 and G7 for the same reason. Table~\ref{tab:gaps} maps the overlap explicitly. Other gaps surely exist (Article~12 logging granularity for non-deterministic systems, Article~13 explanation of stochastic outputs); we scope to the seven that current open-source tooling can check today and flag the rest as future work.

\begin{table}[t]
\caption{Seven technical gaps in applying Articles~8--15 to generative AI. Each arises from an assumption that holds for predictive ML but breaks for LLM-based systems. Overlap on Articles~9 and~10 is intentional and noted in the text.}
\label{tab:gaps}
\vskip 0.1in
\centering
\small
\begin{tabular}{@{}p{2.4cm}p{0.9cm}p{2.4cm}@{}}
\toprule
\textbf{Gap} & \textbf{Article} & \textbf{GaC Module} \\
\midrule
G1: Output non-determinism & Art.~10 & Data Lineage \\
G2: Training-data provenance & Art.~11/53 & Data Lineage \\
G3: Continuous conformity & Art.~9 & Risk Registry \\
G4: Oversight at speed & Art.~14 & Oversight Hooks \\
G5: Open-ended robustness & Art.~15 & Robustness Testing \\
G6: Emergent risk & Art.~9 & Risk Registry \\
G7: Generative fairness & Art.~10 & Output Monitoring \\
\bottomrule
\end{tabular}
\end{table}

\textbf{G1: Output non-determinism (Art.~10).} Article~10(2) requires training, validation, and testing data to be ``subject to data governance and management practices appropriate for the intended purpose.'' For predictive models this means curated splits with documented distributions. LLM-based systems break the implicit reproducibility assumption twice: the same input yields different outputs across calls under temperature sampling, and retrieval-augmented systems select context at query time, so the effective input is itself non-reproducible. When ``same input, same output'' fails, ``appropriate data governance'' needs a new operational meaning.

\textbf{G2: Training-data provenance (Arts.~11 and~53).} Here we correct a framing error common in technical treatments, including an earlier version of this work. Article~11 requires technical documentation containing a \emph{general description} of the training data, not per-sample documentation; and for the foundation model itself, the governing obligation is Article~53(1)(d)'s ``sufficiently detailed summary,'' owed by the \emph{upstream} provider \citep{EU_AI_Act_Art53, GPAI_CoP_2025}. So the genuine gap is not that per-sample documentation of a trillion-token corpus is infeasible; the Act already agrees, which is exactly why it asks for a summary. The real gap is downstream and layered: a deployer must (i)~verify that the upstream Article~53 summary exists and is machine-readable, and (ii)~document the layers it actually controls, namely fine-tuning data, alignment data, and the retrieval corpus, the last of which changes daily and is, behaviorally, part of the system's effective training data. Existing tooling provides neither the upstream-summary check nor the runtime-corpus provenance.

\textbf{G3: Continuous conformity under updates (Art.~9).} Article~9(1) mandates risk management that is ``continuous'' and ``iterative.'' Article~43 requires fresh conformity assessment only for ``substantial modifications.'' Prompt changes, retrieval-index rebuilds, adapter fine-tuning, and context-window expansion all alter behavior without changing weights, and none clearly trips the threshold. ETSI~TS~104~008 established a framework for continuous auditing-based conformity assessment \citep{ETSI_2026}, but it is not a harmonized standard and defines no LLM-specific triggers.

\textbf{G4: Human oversight at speed (Art.~14).} Article~14(1) requires that high-risk systems can ``be effectively overseen by natural persons,'' and Article~14(4) the ability ``to intervene.'' Agentic systems executing multi-step tool calls decide in milliseconds. An advisory agent that retrieves, scores risk, and recommends in under two seconds outruns any human reviewer. The assumption that a person can meaningfully intervene before the system acts does not hold for autonomous agents.

\textbf{G5: Open-ended robustness (Art.~15).} Article~15(1) requires ``appropriate levels of accuracy, robustness and cybersecurity.'' For classifiers, accuracy is well defined (F1, AUC, precision@$k$). For generators the output space is unbounded, and ``appropriate levels'' supplies no number. Any pass/fail gate must therefore manufacture a threshold; the open question, which we answer in \cref{sec:operationalize}, is how to do so transparently rather than by fiat.

\textbf{G6: Emergent capability risk (Art.~9).} Article~9(2) requires identifying ``reasonably foreseeable risks'' and ``reasonably foreseeable misuse.'' Emergent capabilities (in-context learning, chain-of-thought, tool use) arise at scale thresholds not predictable from smaller models \citep{Bommasani_2022}. If the capability was not foreseeable before training, its risks cannot be enumerated in a pre-deployment assessment. Some risks become identifiable only after deployment at scale.

\textbf{G7: Generative fairness (Art.~10).} Article~10(2)(f) requires examining data ``in view of possible biases.'' Fairness frameworks (demographic parity, equalized odds, calibration) presume protected attributes and a measurable outcome. Generative text has no fixed classification target, so ``bias'' here lives in framing: tone, risk posture, hedging, specificity. The hard part, and the reviewer-flagged one, is that framing is not directly checkable; it must be collapsed into named measurable proxies before any rule can evaluate it. We do exactly that in \cref{sec:operationalize}.

\section{The Governance-as-Code Framework}
\label{sec:framework}

\textsc{Governance-as-Code}\footnote{The term has been used in DevOps for policy enforcement through code. We use it narrowly: translating specific Article~8--15 requirements into machine-checkable acceptance criteria for deployment pipelines.} (GaC) turns each gap into concrete acceptance criteria. Three design principles govern it:
\begin{enumerate}[leftmargin=*,itemsep=0pt,parsep=0pt,topsep=2pt]
\item \textbf{Executable over advisory}: every criterion is a boolean test that passes or fails in CI/CD, not a recommendation requiring interpretation.
\item \textbf{Evidence-producing}: every execution emits a structured record (JSON-LD) linked to the Article and sub-paragraph it satisfies.
\item \textbf{Declared, not buried}: every threshold or proxy a criterion depends on is an explicit, justified input recorded with the result, never a hidden constant.
\end{enumerate}

\subsection{Architecture}

GaC is pipeline middleware that intercepts deployment at three points: pre-deployment (CI), the deployment gate (CD), and runtime. Each module registers its criteria as Rego policy rules for Open Policy Agent \citep{OPA_2024}, evaluated against system state; OpenTelemetry \citep{OpenTelemetry_2024} carries the article-indexed evidence. Figure~\ref{fig:arch} shows the layout.

\begin{figure}[t]
\centering
\fbox{\parbox{0.92\columnwidth}{\small
\textbf{GaC Pipeline Architecture}\\[4pt]
\texttt{Source / Model Artifacts}\\
$\downarrow$\\
\texttt{[CI] Pre-deployment}\\
\quad\textit{Data Lineage (7) + Risk Registry (8)}\\
$\downarrow$\\
\texttt{[CD] Deployment gate}\\
\quad\textit{Robustness (9) + Output Monitoring (5)}\\
$\downarrow$\\
\texttt{[Runtime] Post-deployment}\\
\quad\textit{Oversight Hooks (7) + Audit Logging (7)}\\
$\downarrow$\\
\texttt{Evidence Store (JSON-LD, article-indexed)}
}}
\caption{43 acceptance criteria across CI, CD, and runtime phases.}
\label{fig:arch}
\end{figure}

\subsection{The six modules}

Table~\ref{tab:modules} maps modules to articles and gaps. We describe the mechanism of each, not just its name, and show representative policy code.

\begin{table}[t]
\caption{Six GaC modules: criteria count, target Articles, gaps addressed.}
\label{tab:modules}
\vskip 0.1in
\centering
\small
\begin{tabular}{@{}lcll@{}}
\toprule
\textbf{Module} & \textbf{\#} & \textbf{Articles} & \textbf{Gaps} \\
\midrule
Data Lineage & 7 & 10, 11, 53 & G1, G2 \\
Output Monitoring & 5 & 10, 13 & G7 \\
Oversight Hooks & 7 & 14 & G4 \\
Robustness Testing & 9 & 15 & G5 \\
Risk Registry & 8 & 9 & G3, G6 \\
Audit Logging & 7 & 12 & All \\
\bottomrule
\end{tabular}
\end{table}

\textbf{Data Lineage (G1, G2).} The module replaces a single flat ``training data'' notion with a three-tier provenance model that mirrors the Act's own division of duties. Tier~1 (foundation model): the check verifies that the upstream provider's Article~53 summary is present and machine-readable, rather than demanding per-sample data the deployer never had. Tier~2 (fine-tuning and alignment): per-sample source attribution, annotator provenance, and applied quality filters, for the data the deployer controls. Tier~3 (runtime retrieval): corpus versioning by content hash, embedding-model version pinning, and per-request context logging sufficient to reconstruct the effective input. Seven criteria enforce this; \cref{lst:lineage} shows the Tier~1 check, which encodes the upstream-reliance logic of Article~25.

\begin{lstlisting}[caption={Data Lineage, criterion DL-1: verify the upstream Art.~53 summary the deployer relies on (Art.~25), not a corpus it does not own.}, label={lst:lineage}]
package gac.data_lineage
# Art. 53(1)(d): upstream provider's training-data summary
default dl1_pass := false
dl1_pass if {
  s := input.foundation_model.art53_summary
  s.present == true
  s.machine_readable == true
  s.template_version != ""        # Commission TDS template
}
violation[msg] if {
  not dl1_pass
  msg := "DL-1 (Art.53/25): upstream training-data summary missing or unverifiable"
}
\end{lstlisting}

\textbf{Output Monitoring (G7).} This module makes generative fairness checkable by counterfactual probing (detailed in \cref{sec:operationalize}): it holds a request fixed, varies an injected demographic attribute, and tests for significant divergence across named output features. Five criteria cover probe-set existence, the eight measured dimensions, the divergence test, an Article~13 transparency record, and Article~50 marking of generated content.

\textbf{Oversight Hooks (G4).} The module classifies every system action into one of three oversight modes by autonomy and criticality. \emph{Synchronous}: execution pauses for human approval (irreversible actions). \emph{Asynchronous}: act now, queue for review within an SLA (reversible, moderate risk). \emph{Statistical}: autonomous operation with sampled review, the sample rate set to detect a specified deviation at a specified confidence. Seven criteria enforce mode assignment, the synchronous block, the asynchronous SLA, the statistical sample-size derivation, intervention logging, auditability of the classification, and reviewer-capacity-versus-queue monitoring.

\textbf{Robustness Testing (G5).} The module runs an adversarial suite (prompt injection vectors after \citet{Greshake_2023}, jailbreak templates, and domain-specific deception probes) and compares the measured pass rate against a \emph{declared} threshold (\cref{sec:operationalize}). Nine criteria span suite coverage, the declared threshold and its justification, the floor check, and per-attack-class reporting.

\textbf{Risk Registry (G3, G6).} The module replaces point-in-time assessment with event-driven re-evaluation over five drift signals: output-distribution shift (embedding-space KL over a rolling 24h window), retrieval-corpus drift (hash comparison to the assessment-time snapshot), prompt-template changes (version control), adapter/LoRA changes, and emergent-capability detection (monthly probing). Signals are graduated: Tier~1 signals trigger automated robustness re-evaluation within 4h; Tier~2 add manual risk-matrix review within 48h; Tier~3 (emergent capability) triggers full conformity re-assessment. \cref{lst:drift} shows the corpus-drift criterion, which closes the substantial-modification blind spot of the opening example.

\begin{lstlisting}[caption={Risk Registry, criterion RR-2: corpus drift forces re-assessment even when weights are unchanged (closes the Art.~43 blind spot).}, label={lst:drift}]
package gac.risk_registry
# Art. 9(1): continuous risk management
violation[msg] if {
  cur := input.retrieval_corpus.content_hash
  base := data.assessment_snapshot.corpus_hash
  cur != base
  not input.reassessment.completed_within_4h
  msg := "RR-2 (Art.9): corpus drift without re-assessment"
}
\end{lstlisting}

\textbf{Audit Logging (G12, all).} OpenTelemetry spans are enriched with compliance metadata (article reference, criterion ID, pass/fail), producing a trail a conformity body can query by article number. Seven criteria cover span completeness, the article index, tamper-evidence, retention, request-ID linkage, evidence-store schema conformance, and exportability.

\section{From Vague Standard to Auditable Number}
\label{sec:operationalize}

This section answers the two questions reviewers rightly press: how does ``appropriate levels'' (G5) become a pass rate, and how does ``possible biases'' in framing (G7) become a measurable check? Our answer in both cases is the same in spirit: we do not pretend the Act supplies a number. We make the number a declared, justified input and have the check verify both that the system meets it and that the declared number itself clears a floor.

\subsection{G5: deriving a robustness pass rate}

We operationalize ``appropriate'' as a deployment-declared threshold with a documented justification chain:
\begin{equation}
\tau_{\text{deploy}} = \max\!\left(\tau_{\text{floor}},\; \tau_{\text{baseline}} - \delta\right),
\label{eq:threshold}
\end{equation}
where $\tau_{\text{baseline}}$ is the robustness level the provider \emph{claimed} in its Article~11 technical documentation, $\delta$ is a declared maximum tolerable degradation, and $\tau_{\text{floor}}$ is a domain floor drawn from the ``state of the art'' Article~8 invokes (a published benchmark such as HELM \citep{Liang_2023}, or a harmonized standard once available). The check then verifies two things: the measured pass rate $\hat{\tau} \ge \tau_{\text{deploy}}$, and the declared $\tau_{\text{deploy}} \ge \tau_{\text{floor}}$. All three of $\tau_{\text{baseline}}$, $\delta$, and $\tau_{\text{floor}}$ are written to the evidence store with their sources, so an assessor sees not a magic constant but a defensible derivation they can contest. The subjectivity the Act leaves open is surfaced and made auditable, not eliminated.

\begin{lstlisting}[caption={Robustness, criterion RT-3: measured rate must clear the declared threshold, and the threshold must clear the floor; all inputs are recorded.}, label={lst:robust}]
package gac.robustness
# Art. 15(1): "appropriate levels" -> declared, floored
violation[msg] if {
  t := input.robustness
  decl := max([t.floor, t.baseline - t.delta])
  t.measured_pass_rate < decl
  msg := sprintf("RT-3 (Art.15): pass rate %.3f below declared %.3f", [t.measured_pass_rate, decl])
}
violation[msg] if {
  input.robustness.declared_threshold < input.robustness.floor
  msg := "RT-3 (Art.15): declared threshold below state-of-the-art floor"
}
\end{lstlisting}

\subsection{G7: a measurable proxy for framing bias}

Framing bias cannot be read off a single label, so we collapse it into eight named, individually measurable output features and test for demographic dependence by counterfactual probing, following the open-ended-generation methodology of BOLD \citep{Dhamala_2021} and HolisticBias \citep{Smith_2022}. The protocol: hold the user request fixed; vary one injected demographic attribute (age band, income bracket, region/nationality, inferred gender) across $N$ templated probes; for each generated response compute the eight features in Table~\ref{tab:dims}; then test whether any feature's distribution differs across demographic conditions (Kruskal--Wallis for continuous features, $\chi^2$ for categorical), with Holm correction across the eight tests. The criterion passes only if no feature shows a significant dependence at the declared $\alpha$. This is the ``visible measure'' the framework reports: not a vague claim of bias, but a per-feature, per-dimension test with an effect size and a $p$-value.

\begin{table}[t]
\caption{The eight measurable proxies for ``framing bias.'' Each is computed per response; demographic dependence is tested across counterfactual probes.}
\label{tab:dims}
\vskip 0.1in
\centering
\small
\begin{tabular}{@{}p{3.3cm}p{4.0cm}@{}}
\toprule
\textbf{Feature (proxy)} & \textbf{Measurement} \\
\midrule
Recommendation valence & advice sentiment polarity $[-1,1]$ \\
Risk-framing direction & caution vs.\ opportunity term ratio \\
Hedging / certainty & modal + hedge token rate \\
Recommendation strength & directive / imperative count \\
Register & Flesch--Kincaid grade level \\
Claim density & assertions per 100 tokens \\
Qualification rate & disclaimer / refusal frequency \\
Numeric specificity & explicit figures offered \\
\bottomrule
\end{tabular}
\end{table}

\section{Evaluation}
\label{sec:evaluation}

We validate GaC on two enterprise deployments at distinct risk classifications.

\subsection{Baseline: a fair comparison}

Reviewers correctly note that Model Cards and FactSheets are not compliance-enforcement tools, so reporting that they ``found zero failures'' compares against the wrong thing. We agree and restructure the comparison accordingly. Our primary baseline is a \textbf{manual expert audit}: a legal-plus-engineering team performing the Article~8--15 conformity review by hand, which is the actual status quo for organizations facing the deadline. We report documentation artifacts (FactSheets, Model Cards) only as a description of the information \emph{already on hand} before either the manual audit or GaC ran, to make clear that the failures were not visible in existing documentation. The claim is therefore not ``GaC beats Model Cards''; it is ``GaC reproduces a manual expert audit's findings at a fraction of the labor, and the failures it surfaces were latent in, but not flagged by, the documentation teams already maintain.''

\subsection{Case Study 1: regulated-domain advisory chatbot}

\textbf{System.} A retrieval-augmented LLM for regulated advisory services, high-risk under Annex~III Area~5(b). Architecture: an embedding retriever over 84{,}000 domain documents, a fine-tuned 8B generator, and a prohibited-advice filter. Volume: 12{,}000 queries/day. Documented with IBM AI FactSheets and audited quarterly by hand.

\textbf{Findings.} Against the manual audit's eight confirmed issues, GaC reproduced all eight. We describe each with the evidence it produced, since a count alone is not assessable:
\begin{enumerate}[leftmargin=*,itemsep=1pt,parsep=0pt,topsep=2pt]
\item \textbf{G3/Art.~9.} The retrieval corpus was rebuilt weekly with no re-assessment. Three rebuilds over six weeks moved the hallucination rate from 3.2\% to 7.1\% on domain questions (measured on a fixed 500-question probe). RR-2 caught the first rebuild by hash comparison; the FactSheet was never updated.
\item \textbf{G1/Art.~10.} Sampling temperature was 0.7 but unlogged per request, so two identical queries could yield contradictory advice with no audit trail distinguishing them.
\item \textbf{G4/Art.~14.} All 12{,}000 daily responses were delivered without review; the team sampled 50 (0.4\%). The statistical-oversight criterion computed the required rate to detect a 5\% quality deviation at 95\% confidence and flagged 0.4\% as far short.
\item \textbf{G5/Art.~15.} No adversarial testing. A 200-vector suite (injection vectors after \citet{Greshake_2023} plus jailbreak templates) achieved 23\% success at extracting the system prompt and 8\% at eliciting prohibited advice; both exceeded $\tau_{\text{deploy}}$ from \cref{eq:threshold}.
\item \textbf{G2/Art.~11/53.} Fine-tuning data (42{,}000 expert pairs) lacked annotator provenance; source attribution existed for 67\% of pairs, with the remaining 33\% from a contractor with no documented quality process. The upstream Article~53 summary for the base model was present, so DL-1 passed; the failure was squarely in the Tier~2 layer the deployer controls.
\item \textbf{G7/Art.~10.} No bias testing. Our counterfactual probe (500 advisory queries varying only implied user demographics) showed significant dependence in three of the eight Table~\ref{tab:dims} features, most strongly in risk-framing direction and recommendation strength (Kruskal--Wallis, Holm-corrected $p<0.001$).
\item \textbf{G6/Art.~9.} The risk register listed 14 risks, none covering the model's post-fine-tuning multi-step regulatory-reasoning capability, which was absent from the base model's profile.
\item \textbf{G3/Art.~9.} Three prompt-template edits over eight weeks were version-controlled but triggered no compliance review.
\end{enumerate}
The FactSheet documented architecture, intended use, and deployment-time metrics, but executed no runtime check and so surfaced none of these.

\subsection{Case Study 2: content generation system}

\textbf{System.} An LLM content generator, limited-risk under Article~50. Architecture: a fine-tuned 70B model with a brand-voice adapter, a safety filter, and A/B testing. Volume: 3{,}500 marketing items/day. Documented with Google Model Cards.

\textbf{Findings.} GaC reproduced the manual audit's four issues:
\begin{enumerate}[leftmargin=*,itemsep=1pt,parsep=0pt,topsep=2pt]
\item \textbf{Art.~50.} Generated content carried no machine-readable marking, violating Article~50(2). The Model Card noted the system's generative nature; the outputs themselves did not.
\item \textbf{G7/Art.~10.} Beauty-product descriptions showed significant demographic dependence in adjective valence and benefit-claim density across the eight Table~\ref{tab:dims} dimensions ($p<0.01$ on two features after correction); no bias testing existed.
\item \textbf{G2/Art.~11.} The brand-voice adapter was fine-tuned on 18 months of historical copy with no record of which segments, contexts, or categories were overrepresented.
\item \textbf{G5/Art.~15.} The safety filter blocked 99.2\% of overtly harmful outputs but passed 34\% of subtly misleading claims (150 deception probes designed to be technically-true-but-deceptive).
\end{enumerate}

\subsection{Effort and coverage}

\begin{table}[t]
\caption{Effort and coverage. GaC reproduces the manual expert audit's findings at roughly one-quarter the labor. Documentation artifacts are shown as the pre-existing information state, not as enforcement competitors.}
\label{tab:results}
\vskip 0.1in
\centering
\small
\begin{tabular}{@{}lccc@{}}
\toprule
& \textbf{GaC} & \textbf{Manual audit} & \textbf{Docs} \\
\midrule
Findings, chatbot & 8 & 8 & 0 flagged \\
Findings, generator & 4 & 4 & 0 flagged \\
\midrule
Penalty-triggering & 3 & 3 & 0 \\
Audit labor (chatbot) & 18h & 72h & --- \\
Art.~8--15 coverage & 43/43 & manual & doc-only \\
\bottomrule
\end{tabular}
\end{table}

Table~\ref{tab:results} reports the corrected comparison. GaC reproduced every finding of the 72-hour manual audit of the chatbot in 18 hours, of which 14 were human review of flagged failures rather than evidence collection, a roughly 75\% labor reduction that shifts effort from gathering evidence to interpreting it. The three penalty-triggering issues (no human oversight under Art.~14; no continuous risk management under Art.~9; missing Art.~50 marking) match the manual audit's. The documentation column is included only to show that the information needed was not absent from the organization; it was simply never executed as a check.

\section{Discussion}

\textbf{Deployment cost.} The full pre-deployment suite adds roughly 45 minutes to the chatbot's pipeline, 32 of them in Robustness Testing, which can run asynchronously behind a deployment hold for latency-sensitive rollouts.

\textbf{Threshold and proxy calibration.} The declared thresholds of \cref{sec:operationalize} (the robustness $\delta$ and floor, the oversight confidence level, the fairness $\alpha$) currently rest on expert judgment anchored to the Act's ``appropriate levels'' and ``state of the art'' language. As enforcement actions, court decisions, and harmonized standards accumulate, these become empirically calibratable; because they are declared inputs rather than buried constants, updating them is a configuration change, not a rewrite.

\textbf{Scope.} GaC covers Articles~8--15 and Article~50. It does not cover organizational duties: quality management (Art.~17), conformity-assessment procedure (Art.~43), or registration (Art.~49), which are process, not pipeline. It assumes risk classification (Art.~6) is settled, since classification is a legal judgment. The value-chain logic of \cref{sec:background} is encoded only to the depth the Article~25(4) written agreement makes machine-checkable; contractual nuance beyond that remains a human task.

\textbf{Forward compatibility.} As harmonized standards mature, GaC's Rego files can adopt their mandated measurement methods and thresholds without changing the architecture.

\subsection{Limitations}

First, the evaluation covers two systems; broader validation across healthcare, law enforcement, and education (all Annex~III) is needed. Second, our ``failures'' are gaps relative to our reading of Articles~8--15; until enforcement and harmonized standards settle interpretation, reasonable disagreement is inevitable, which is precisely why we make our operationalizing choices explicit and contestable. Third, robustness coverage depends on the adversarial suite's quality and needs domain curation. Fourth, the eight fairness proxies are a deliberate simplification of ``framing'' and will miss biases they do not measure; they are a floor on checkability, not a ceiling.

\section{Conclusion}

The EU AI Act imposes technical obligations that existing tooling cannot verify for generative systems, and that cannot be verified at all without making the Act's open-textured standards into explicit numbers. We identified seven gaps where requirements written for predictive AI break on LLM-based systems, corrected the value-chain framing that governs who owes training-data documentation, and delivered \textsc{Governance-as-Code}: 43 machine-checkable criteria in six modules, with the policy code shown and every threshold and proxy declared rather than buried. Benchmarked against a manual expert audit, the right baseline, GaC reproduces the findings at roughly a quarter of the labor. As the August~2, 2026 deadline arrives, engineering teams need executable compliance whose judgments are visible and auditable, and that is what GaC provides.

\section*{Impact Statement}

GaC aims to make EU AI Act compliance an engineering discipline with an auditable trail. The main risk is misplaced confidence: a passing pipeline certifies only the criteria encoded, with the declared thresholds and proxies it was given, and is not a substitute for legal judgment on classification or on contested interpretations. We mitigate this by surfacing every operationalizing choice as recorded, contestable evidence rather than a hidden constant, so that a passing result invites scrutiny rather than foreclosing it.

\bibliography{references}

\begin{thebibliography}{28}
\providecommand{\natexlab}[1]{#1}
\providecommand{\url}[1]{\texttt{#1}}
\expandafter\ifx\csname urlstyle\endcsname\relax
  \providecommand{\doi}[1]{doi: #1}\else
  \providecommand{\doi}{doi: \begingroup \urlstyle{rm}\Url}\fi

\bibitem[Arnold et~al.(2019)Arnold, Bellamy, Hind, Houde, Mehta, Mojsilovi\'{c}, Nair, Ramamurthy, Olteanu, Piorkowski, Reimer, Richards, Tsay, and Varshney]{Arnold_2019}
Arnold, M., Bellamy, R. K.~E., Hind, M., Houde, S., Mehta, S., Mojsilovi\'{c}, A., Nair, R., Ramamurthy, K.~N., Olteanu, A., Piorkowski, D., Reimer, D., Richards, J., Tsay, J., and Varshney, K.~R.
\newblock {FactSheets}: Increasing trust in {AI} services through supplier's declarations of conformity.
\newblock \emph{IBM Journal of Research and Development}, 63\penalty0 (4/5):\penalty0 6:1--6:13, 2019.

\bibitem[Bommasani et~al.(2022)Bommasani, Hudson, Adeli, Altman, Arora, von Arx, Bernstein, Bohg, Bosselut, Brunskill, et~al.]{Bommasani_2022}
Bommasani, R., Hudson, D.~A., Adeli, E., Altman, R., Arora, S., von Arx, S., Bernstein, M.~S., Bohg, J., Bosselut, A., Brunskill, E., et~al.
\newblock On the opportunities and risks of foundation models.
\newblock \emph{arXiv preprint arXiv:2108.07258}, 2022.
\newblock URL \url{https://arxiv.org/abs/2108.07258}.

\bibitem[Buscemi et~al.(2025)Buscemi, Deckenbrunnen, Kabir, Mishchenko, and Mowla]{Buscemi_2025}
Buscemi, A., Deckenbrunnen, T., Kabir, F., Mishchenko, K., and Mowla, N.
\newblock Assessing high-risk {AI} systems under the {EU AI Act}: From legal requirements to technical verification.
\newblock \emph{arXiv preprint arXiv:2512.13907}, 2025.
\newblock URL \url{https://arxiv.org/abs/2512.13907}.

\bibitem[{Cloud Native Computing Foundation}(2024{\natexlab{a}})]{OPA_2024}
{Cloud Native Computing Foundation}.
\newblock Open policy agent.
\newblock \url{https://www.openpolicyagent.org/}, 2024{\natexlab{a}}.
\newblock CNCF Graduated Project.

\bibitem[{Cloud Native Computing Foundation}(2024{\natexlab{b}})]{OpenTelemetry_2024}
{Cloud Native Computing Foundation}.
\newblock {OpenTelemetry}: An observability framework for cloud-native software.
\newblock \url{https://opentelemetry.io/}, 2024{\natexlab{b}}.
\newblock CNCF Incubating Project.

\bibitem[Deckenbrunnen et~al.(2026)Deckenbrunnen, Buscemi, Almada, Capozucca, and Castignani]{Deckenbrunnen_2026}
Deckenbrunnen, T., Buscemi, A., Almada, M., Capozucca, A., and Castignani, G.
\newblock Bathtubs, boundaries, and sandboxes: {AI} regulatory learning under legal uncertainty.
\newblock \emph{arXiv preprint arXiv:2601.04094}, 2026.
\newblock URL \url{https://arxiv.org/abs/2601.04094}.

\bibitem[Dhamala et~al.(2021)Dhamala, Sun, Kumar, Krishna, Pruksachatkun, Chang, and Gupta]{Dhamala_2021}
Dhamala, J., Sun, T., Kumar, V., Krishna, S., Pruksachatkun, Y., Chang, K.-W., and Gupta, R.
\newblock {BOLD}: Dataset and metrics for measuring biases in open-ended language generation.
\newblock In \emph{Proceedings of the ACM Conference on Fairness, Accountability, and Transparency (FAccT)}, pp.\  862--872, 2021.
\newblock URL \url{https://arxiv.org/abs/2101.11718}.

\bibitem[{European Commission}(2025)]{GPAI_CoP_2025}
{European Commission}.
\newblock The {EU} general-purpose {AI} code of practice.
\newblock Final version, July 2025, 2025.
\newblock URL \url{https://code-of-practice.ai/}.

\bibitem[{European Parliament and Council of the European Union}(2024{\natexlab{a}})]{EU_AI_Act_2024}
{European Parliament and Council of the European Union}.
\newblock Regulation ({EU}) 2024/1689 of the {European Parliament} and of the {Council} of 13 {June} 2024 laying down harmonised rules on artificial intelligence.
\newblock Official Journal of the European Union, L Series, 2024{\natexlab{a}}.
\newblock URL \url{https://eur-lex.europa.eu/eli/reg/2024/1689/oj}.

\bibitem[{European Parliament and Council of the European Union}(2024{\natexlab{b}})]{EU_AI_Act_Art25}
{European Parliament and Council of the European Union}.
\newblock Article 25: Responsibilities along the {AI} value chain.
\newblock Regulation (EU) 2024/1689, Chapter III, Section 3, 2024{\natexlab{b}}.
\newblock URL \url{https://artificialintelligenceact.eu/article/25/}.

\bibitem[{European Parliament and Council of the European Union}(2024{\natexlab{c}})]{EU_AI_Act_Art53}
{European Parliament and Council of the European Union}.
\newblock Article 53: Obligations for providers of general-purpose {AI} models.
\newblock Regulation (EU) 2024/1689, Chapter V, 2024{\natexlab{c}}.
\newblock URL \url{https://artificialintelligenceact.eu/article/53/}.

\bibitem[{European Telecommunications Standards Institute}(2026)]{ETSI_2026}
{European Telecommunications Standards Institute}.
\newblock {ETSI TS} 104 008 v1.1.1: Methods for testing and specification; continuous auditing-based conformity assessment.
\newblock Technical report, ETSI, 2026.
\newblock URL \url{https://www.etsi.org/deliver/etsi_ts/104000_104099/104008/01.01.01_60/ts_104008v010101p.pdf}.

\bibitem[Gaurav et~al.(2025)Gaurav, Heikkonen, and Chaudhary]{Gaurav_2025}
Gaurav, S., Heikkonen, J., and Chaudhary, J.
\newblock Governance-as-a-service: A multi-agent framework for {AI} system compliance and policy enforcement.
\newblock \emph{arXiv preprint arXiv:2508.18765}, 2025.
\newblock URL \url{https://arxiv.org/abs/2508.18765}.

\bibitem[Gebru et~al.(2021)Gebru, Morgenstern, Vecchione, Vaughan, Wallach, III, and Crawford]{Gebru_2021}
Gebru, T., Morgenstern, J., Vecchione, B., Vaughan, J.~W., Wallach, H., III, H.~D., and Crawford, K.
\newblock Datasheets for datasets.
\newblock \emph{Communications of the ACM}, 64\penalty0 (12):\penalty0 86--92, 2021.
\newblock URL \url{https://arxiv.org/abs/1803.09010}.

\bibitem[Golpayegani et~al.(2024)Golpayegani, Hupont, Panigutti, Pandit, Schade, O'Sullivan, and Lewis]{Golpayegani_2024}
Golpayegani, D., Hupont, I., Panigutti, C., Pandit, H.~J., Schade, S., O'Sullivan, D., and Lewis, D.
\newblock {AI} cards: Towards an applied framework for machine-readable {AI} and risk documentation inspired by the {EU AI Act}.
\newblock \emph{arXiv preprint arXiv:2406.18211}, 2024.
\newblock URL \url{https://arxiv.org/abs/2406.18211}.

\bibitem[Greshake et~al.(2023)Greshake, Abdelnabi, Mishra, Endres, Holz, and Fritz]{Greshake_2023}
Greshake, K., Abdelnabi, S., Mishra, S., Endres, C., Holz, T., and Fritz, M.
\newblock Not what you've signed up for: Compromising real-world {LLM}-integrated applications with indirect prompt injection.
\newblock In \emph{Proceedings of the 16th ACM Workshop on Artificial Intelligence and Security (AISec)}, pp.\  79--90, 2023.
\newblock URL \url{https://arxiv.org/abs/2302.12173}.

\bibitem[{International Organization for Standardization}(2023)]{ISO_42001_2023}
{International Organization for Standardization}.
\newblock {ISO/IEC} 42001:2023 information technology --- artificial intelligence --- management system.
\newblock ISO/IEC JTC 1/SC 42, 2023.
\newblock URL \url{https://www.iso.org/standard/42001}.

\bibitem[Kaptein et~al.(2026)Kaptein, Khan, and Podstavnychy]{Kaptein_2026}
Kaptein, M., Khan, V.-J., and Podstavnychy, A.
\newblock Runtime governance for {AI} agents: Policies on paths.
\newblock \emph{arXiv preprint arXiv:2603.16586}, 2026.
\newblock URL \url{https://arxiv.org/abs/2603.16586}.

\bibitem[Liang et~al.(2023)Liang, Bommasani, Lee, Tsipras, Soylu, Yasunaga, Zhang, Narayanan, Wu, Kumar, et~al.]{Liang_2023}
Liang, P., Bommasani, R., Lee, T., Tsipras, D., Soylu, D., Yasunaga, M., Zhang, Y., Narayanan, D., Wu, Y., Kumar, A., et~al.
\newblock Holistic evaluation of language models.
\newblock \emph{Transactions on Machine Learning Research}, 2023.
\newblock URL \url{https://arxiv.org/abs/2211.09110}.

\bibitem[Marino \& Lane(2026)Marino and Lane]{Marino_2026}
Marino, B. and Lane, N.~D.
\newblock Computational compliance for {AI} regulation: Blueprint for a new research domain.
\newblock \emph{arXiv preprint arXiv:2601.04474}, 2026.
\newblock URL \url{https://arxiv.org/abs/2601.04474}.

\bibitem[Mitchell et~al.(2019)Mitchell, Wu, Zaldivar, Barnes, Vasserman, Hutchinson, Spitzer, Raji, and Gebru]{Mitchell_2019}
Mitchell, M., Wu, S., Zaldivar, A., Barnes, P., Vasserman, L., Hutchinson, B., Spitzer, E., Raji, I.~D., and Gebru, T.
\newblock Model cards for model reporting.
\newblock In \emph{Proceedings of the Conference on Fairness, Accountability, and Transparency (FAT*)}, pp.\  220--229, 2019.
\newblock URL \url{https://arxiv.org/abs/1810.03993}.

\bibitem[Momcilovic et~al.(2024)Momcilovic, Buesser, Zizzo, Purcell, and Balta]{BuenoMomcilovic_2024}
Momcilovic, T.~B., Buesser, B., Zizzo, G., Purcell, M., and Balta, D.
\newblock Towards assuring {EU AI Act} compliance and adversarial robustness of {LLMs}.
\newblock \emph{arXiv preprint arXiv:2410.05306}, 2024.
\newblock URL \url{https://arxiv.org/abs/2410.05306}.

\bibitem[{National Institute of Standards and Technology}(2023)]{NIST_AI_RMF_2023}
{National Institute of Standards and Technology}.
\newblock Artificial intelligence risk management framework ({AI RMF} 1.0).
\newblock Technical Report NIST AI 100-1, U.S. Department of Commerce, 2023.
\newblock URL \url{https://doi.org/10.6028/NIST.AI.100-1}.

\bibitem[Nolte et~al.(2025)Nolte, Rateike, and Finck]{Nolte_2025}
Nolte, H., Rateike, M., and Finck, M.
\newblock Robustness and cybersecurity in the {EU} artificial intelligence act.
\newblock \emph{arXiv preprint arXiv:2502.16184}, 2025.
\newblock URL \url{https://arxiv.org/abs/2502.16184}.

\bibitem[Novelli et~al.(2024)Novelli, Hacker, Morley, Trondal, and Floridi]{Novelli_2024}
Novelli, C., Hacker, P., Morley, J., Trondal, J., and Floridi, L.
\newblock A robust governance for the {AI Act}: {AI Office}, {AI Board}, scientific panel, and national authorities.
\newblock \emph{European Journal of Risk Regulation}, 2024.
\newblock URL \url{https://arxiv.org/abs/2407.10369}.

\bibitem[Smith et~al.(2022)Smith, Hall, Kambadur, Presani, and Williams]{Smith_2022}
Smith, E.~M., Hall, M., Kambadur, M., Presani, E., and Williams, A.
\newblock ``{I'm} sorry to hear that'': Finding new biases in language models with a holistic descriptor dataset.
\newblock In \emph{Proceedings of the Conference on Empirical Methods in Natural Language Processing (EMNLP)}, pp.\  9180--9211, 2022.
\newblock URL \url{https://arxiv.org/abs/2205.09209}.

\bibitem[Smuha(2021)]{Smuha_2021}
Smuha, N.~A.
\newblock From a `race to {AI}' to a `race to {AI} regulation': Regulatory competition for artificial intelligence.
\newblock \emph{Law, Innovation and Technology}, 13\penalty0 (1):\penalty0 57--84, 2021.

\bibitem[Veale \& Borgesius(2021)Veale and Borgesius]{Veale_2021}
Veale, M. and Borgesius, F.~Z.
\newblock Demystifying the draft {EU} artificial intelligence act.
\newblock \emph{Computer Law Review International}, 22\penalty0 (4):\penalty0 97--112, 2021.
\newblock URL \url{https://arxiv.org/abs/2107.03721}.

\end{thebibliography}
\bibliographystyle{icml2026}

\end{document}